\documentclass[conference]{IEEEtran}

\usepackage{amsmath,amssymb, graphicx}
\usepackage[ruled,linesnumbered]{algorithm2e}
\usepackage{algorithmic}
\usepackage{paralist} 
\usepackage{color}
\usepackage{multirow}
\usepackage{rotating}
\usepackage{graphicx,wrapfig,lipsum}
\usepackage{hyperref}
\usepackage[mathscr]{eucal} 
\usepackage{amsbsy} 
\usepackage{subfigure}
\usepackage{booktabs}
\usepackage{xcolor}
\usepackage{tikz}
\usetikzlibrary{snakes}
\usepackage[outercaption]{sidecap}  
\usepackage{multirow}
\usepackage{color}
\usepackage{epstopdf}
\usepackage{balance}
\usepackage{tablefootnote}
\usepackage{empheq} 
\usepackage{moresize}
\usepackage{graphicx}

\usepackage{enumitem}
\setlist{nolistsep}
\usepackage{mdframed}
\usepackage{stackengine} 
\usepackage{sidecap}

\usepackage[font=small,labelfont=bf,skip=0pt]{caption}
\DeclareCaptionType{copyrightbox}
\DeclareCaptionType{Algorithm}
\usepackage{url}

\newcounter{ALC@tempcntr}
\newcommand{\mourmeth}{\text{CPR}}
\newcommand{\newmethod}{$\mourmeth$\xspace}

\newcommand{\hide}[1]{}

\newcommand{\bit}{\begin{compactitem}}
\newcommand{\eit}{\end{compactitem}}
\newcommand{\ben}{\begin{compactenum}}
\newcommand{\een}{\end{compactenum}}

\begin{document}

\title{\newmethod: Leveraging LLMs for Topic and Phrase Suggestion to Facilitate Comprehensive Product Reviews}
\author{
\IEEEauthorblockN{Ekta Gujral}
\IEEEauthorblockA{\textit{Walmart Global Technology} \\
ekta.gujral@walmart.com}
\and 
\IEEEauthorblockN{Apurva Sinha}
\IEEEauthorblockA{\textit{Walmart Global Technology} \\
apurva.sinha@walmart.com}
\and
\IEEEauthorblockN{Lishi Ji\textsuperscript}
\IEEEauthorblockA{\textit{Walmart Global Technology} \\
Lishi.Ji@walmart.com}
\and
\IEEEauthorblockN{Bijayani Sanghamitra Mishra}
\IEEEauthorblockA{\textit{Walmart Global Technology} \\
Bijayani.Mishra@walmart.com}
}

\maketitle

\begin{abstract}
Consumers often heavily rely on online product reviews, analyzing both quantitative ratings and textual descriptions to assess product quality. However, existing research hasn't adequately addressed how to systematically encourage the creation of comprehensive reviews that capture both customers sentiment and detailed product feature analysis. This paper presents \newmethod, a novel methodology that leverages the power of Large Language Models (LLMs) and Topic Modeling to guide users in crafting insightful and well-rounded reviews. Our approach employs a three-stage process: first, we present users with product-specific terms for rating; second, we generate targeted phrase suggestions based on these ratings; and third, we integrate user-written text through topic modeling, ensuring all key aspects are addressed. We evaluate \newmethod using text-to-text LLMs, comparing its performance against real-world customer reviews from Walmart. Our results demonstrate that \newmethod effectively identifies relevant product terms, even for new products lacking prior reviews, and provides sentiment-aligned phrase suggestions, saving users time and enhancing reviews quality. Quantitative analysis reveals a 12.3\% improvement in BLEU score over baseline methods, further supported by manual evaluation of generated phrases. We conclude by discussing potential extensions and future research directions.
\end{abstract}

\section{Introduction}
\label{sec:intro}
 Product reviews play a crucial role for retailers, as they help build trust among potential customers by providing social proof. They influence purchase decisions \cite{chakraborty2022attribute,chintagunta2010effects,ludwig2013more,Gujralpae2024} by offering information on the quality and suitability of the product. Reviews also provide valuable feedback for retailers, allows them to improve their products and enhance customer satisfaction. Furthermore, product reviews contribute to product search optimization efforts \cite{chevalier2018channels}, giving retailers a competitive advantage and fostering customer engagement and loyalty. 

\hide{
\textit{\textcolor{blue}{Informal Problem 1.} Given a set of target product information, overall ratings and optional unstructured information in the form of text: how can we generate text which summarizes the sentiment of customers? What if customers have multiple sentiments about the purchased product, like positive sentiment for picture quality and neutral sentiment for battery life }
}
\begin{figure}
	\begin{center}
	    \includegraphics[clip,trim=0cm 3cm 0cm 3cm,width=0.45\textwidth]{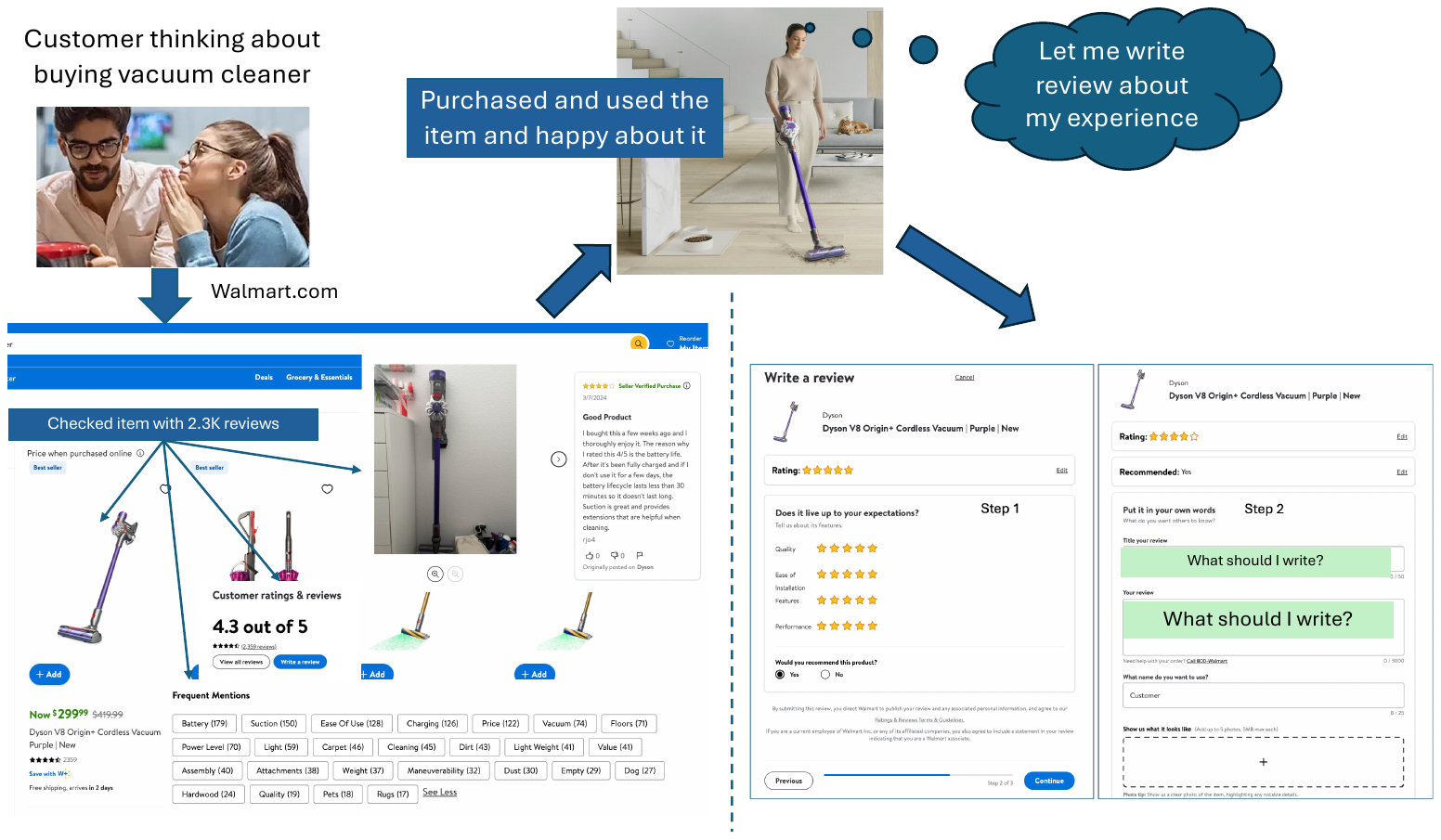}
		\caption{Illustration of Customer's Online Shopping Journey} 
		\label{fig:exampleproblem}
	\end{center}
 \vspace{-0.2in}
\end{figure}
Product review phrase suggestion is a sub-task of text-to-text generation in natural language processing (NLP). Online shopping is increasingly popular. However, customers often lack the motivation to write constructive reviews. At the same time, retailers eagerly seek positive reviews for their own gain. Hence, automatic phrase suggestion allows  retailers to save customers time and effort that would otherwise be spent on manually writing or soliciting reviews. Authentic reviews \cite{zang2017towards,carlson2023complementing,perez2023efficiency,fan2022automatic,tai2018automatic,costa2018automatic} play a crucial role in building trust and credibility among potential customers, influencing their purchasing decisions. Therefore, this can greatly benefit retailers by boosting their reputation, increasing sales, and ultimately driving business growth. 

\textbf{\textcolor{blue}{Motivating Example}}
Consider an example of a customer looking to purchase a new vacuum cleaner for their newly renovated home. They were overwhelmed by the multitude of options available and wanted to make sure that they were investing in a high-quality product. The customer decided to go to the Walmart website to research different vacuum cleaners. They found a particular model that caught their attention due to its affordable price and positive reviews. The vacuum had an average rating of $4.5$ stars out of $5$, with numerous customers praising its powerful suction, durability, and ease of use. Intrigued by the positive feedback, the customer decided to purchase the vacuum cleaner based on the reviews they read as shown in Figure \ref{fig:exampleproblem}. They were excited to receive their new purchase and put it to the test. Once the vacuum arrived, the customer was pleasantly surprised by its performance. The suction power exceeded their expectations, effortlessly picking up dirt, pet hair, and debris from their carpets and hardwood floors. The durability of the vacuum also proved to be excellent, as it withstood regular use without any issues. The positive experience with the vacuum cleaner reinforced the trust of the customer in the reviews on Walmart. They realized that relying on the feedback and experiences of other customers, they were able to make an informed decision and find a high-quality product that perfectly suited their needs. Inspired by this positive experience, the customer continued to explore and trust the reviews at Walmart for future purchases. They found that by doing so they were able to save time and money by investing in products that had been tested and tested by other customers, ultimately leading to their satisfaction and peace of mind. The customer then wanted to give feedback on the product on the website. However, the customer became overwhelmed by the time-consuming review writing process. Here, a topic (ideas) and related phrase suggestion system plays an important role. We can show customers some topics related to the product and ask them to rate them properly. Once the rating is complete, the system can provide 1-2 phrases and topics. The customer can include these in the review. This makes an otherwise overwhelming process simple and efficient. As more and more customers provide authentic and real reviews, retailers will gain trust and drive large sales.

\begin{figure}
	\begin{center}
	    \includegraphics[clip,trim=2cm 3cm 2cm 3cm,width=0.45\textwidth]{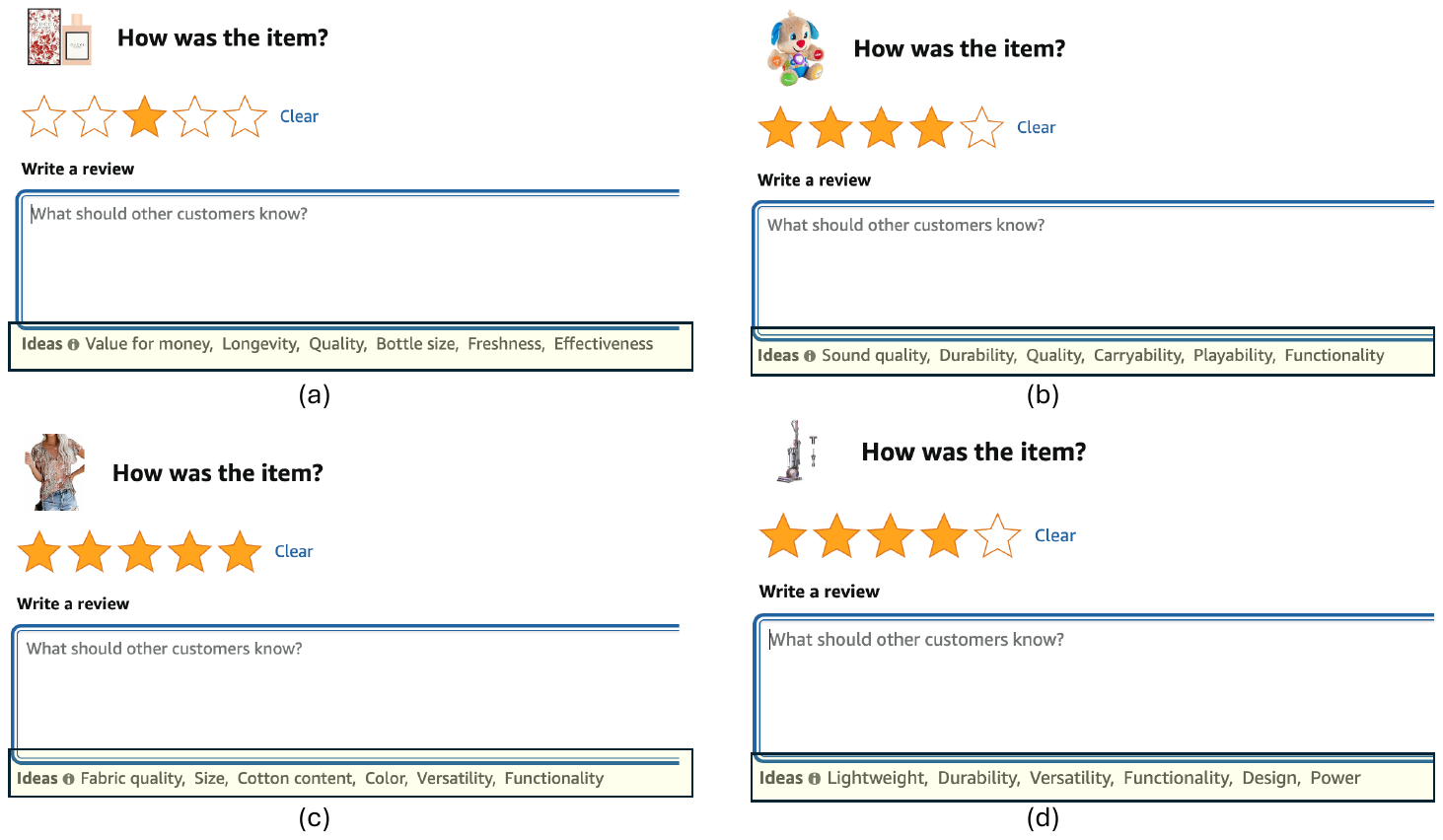}
		\caption{Illustration of other e-commerce mobile platform's product attributes or topics suggestion during review writing process.} 
		\label{fig:amazonexample}
	\end{center}
 \vspace{-0.2in}
\end{figure}

\textbf{\textcolor{blue}{Previous Works on Reviews}}
 Automatic text generation of products has gained attention in recent years, with several studies and research papers exploring this area. Li et al. \cite{korhonen2019proceedings} proposed a method that uses a deep learning model to generate review summaries for products by leveraging user-generated reviews. Their approach utilizes an attention mechanism to capture the most relevant information from the reviews and generate concise and coherent summaries. Similarly, Wang et al. \cite{ruder2016hierarchical} proposed a neural network-based model that generates product reviews by considering both the product features and sentiment information. Their model incorporates a sentiment attention mechanism to ensure the reviews reflect the overall sentiment of the product. Another study by Zhao et al. \cite{zhao2019review} focused on generating review suggestions for products. They proposed a reinforcement learning-based approach that learns to generate coherent and helpful review suggestions by maximizing the reward signal obtained from user feedback. Similarly, Keith et al. \cite{carlson2023complementing} proposed approach that utilizes a deep learning model called the Variational Autoencoder (VAE) to learn the latent representation of review data. To generate review content, the system takes user inputs such as a product description or features, and maps them to the learned latent space using the trained VAE. From this latent representation, the model can generate new review content that is semantically similar to the input but different in wording. This process allows the system to automatically generate personalized and relevant review content. The Perez-Castro et al. \cite{perez2023efficiency} studied a variety of deep learning models, such as Variational Autoencoder (VAE) and Recurrent Neural Network (RNN), for review generation task. The authors evaluate the generated reviews against those written by humans, using metrics like the BLEU score, ROUGE score, and perplexity to assess the quality of the generated text. They also take into account the time taken to train the models and generate reviews as a measure of efficiency. Another study done by Yu et al. \cite{tai2018automatic} involves training an RNN model using a large dataset of reviews from the target domain. The model is designed to generate review content based on a given input, such as a product description or user feedback. The authors leverage the sequential nature of reviews to capture the context and generate coherent and meaningful review content. These studies contribute to the development of automatic review generation systems, offering potential applications in e-commerce and other domains. However, existing automatic review generation methods often lack the ability to fully understand the context of a product types, which can lead to the generation of irrelevant or inaccurate reviews. Also, systems may struggle to accurately capture and convey the sentiments of customer. They may not be able to correctly interpret and generate reviews that reflect nuanced emotions or opinions. Also, automatic reviews invite the fraudulent reviews as bots can learn the system and automatically select the generated review and post it. Recently, Other e-commerce mobile platform guide reviewers by suggesting and detect key product attributes or topics to focus when writing the review on as shown in Figure \ref{fig:amazonexample}. These suggestions aim to help customers provide more structured and helpful feedback.  For example, for a product like a vacuum cleaner, it might prompt reviewers to consider factors like suction power, noise level, ease of use, and battery life.  This helps ensure reviews are comprehensive and cover the most important aspects of the product. \textit{Similarly, away from automatic generating the full review, we propose an efficient workflow that provide topics or ideas to discuss in the review and produce reliable phrases (based on the rating to the topics) to selected topics to save our customers valuable time.}

\textbf{\textcolor{blue}{Challenges}}
This endeavor is motivated by the possibility of addressing two particular review-related challenges. First, automatic review generation methods often struggle to generate personalized responses to customer feedback. They usually resort to generic responses that may not fully address the customer's concerns or queries. These methods may struggle to accurately interpret customers' sentiments \cite{zang2017towards}, especially when they are expressed subtly or through sarcasm. Also, system may fail to empathize with the customer's frustration or disappointment, leading to unsatisfactory responses. No matter how advanced, automatic responses can lack the emotional intelligence and personal touch that a human can provide. This can result in a lack of trust or connection with the customers.

The second challenge is linked to the lack of contextual understanding \cite{guo2021conditional,celikyilmaz2020evaluation}. Contextual understanding refers to the ability to understand the meaning of a text based on the surrounding information. It requires a deep understanding of language, cultural nuances, slang, idioms, and other complex aspects of human communication. The lack of contextual understanding often results in reviews that may be grammatically correct but lack depth or fails to convey the necessary information accurately \cite{voss1980text}. For example, an AI-generated review might not understand that a "cool" gadget isn't necessarily cold in temperature. It may also fail to understand the difference between a product being "light-weight" (not heavy) and "light-weight" (not durable). Moreover, if a product has specific technical aspects, the AI might not fully understand or adequately explain those aspects due to lack of contextual understanding. This can lead to less accurate and meaningless reviews. 

Third, the ease of automatic review generation amplifies the potential for widespread fraud. AI tools can rapidly produce vast quantity of deceptive reviews, artificially inflating product rating's or damaging competitor reputations. This manipulation erodes consumer trust and creates an uneven marketplace, where genuine businesses are disadvantaged. Furthermore, regulatory bodies like the FTC \cite{ftcfakereview2023} are increasingly scrutinizing these practices, leading to potential legal repercussions for those who engage in deceptive review generation.

Our proposed method tackles the limitations of generic responses and lack of contextual understanding. It does this by incorporating advanced sentiment analysis. This allows for personalized, nuanced responses, moving beyond simple keyword matching and addressing the complexities of human language.

\begin{figure*}
	\begin{center}
	    \includegraphics[clip,trim=0cm 3cm 0cm 3cm,width=0.65\textwidth]{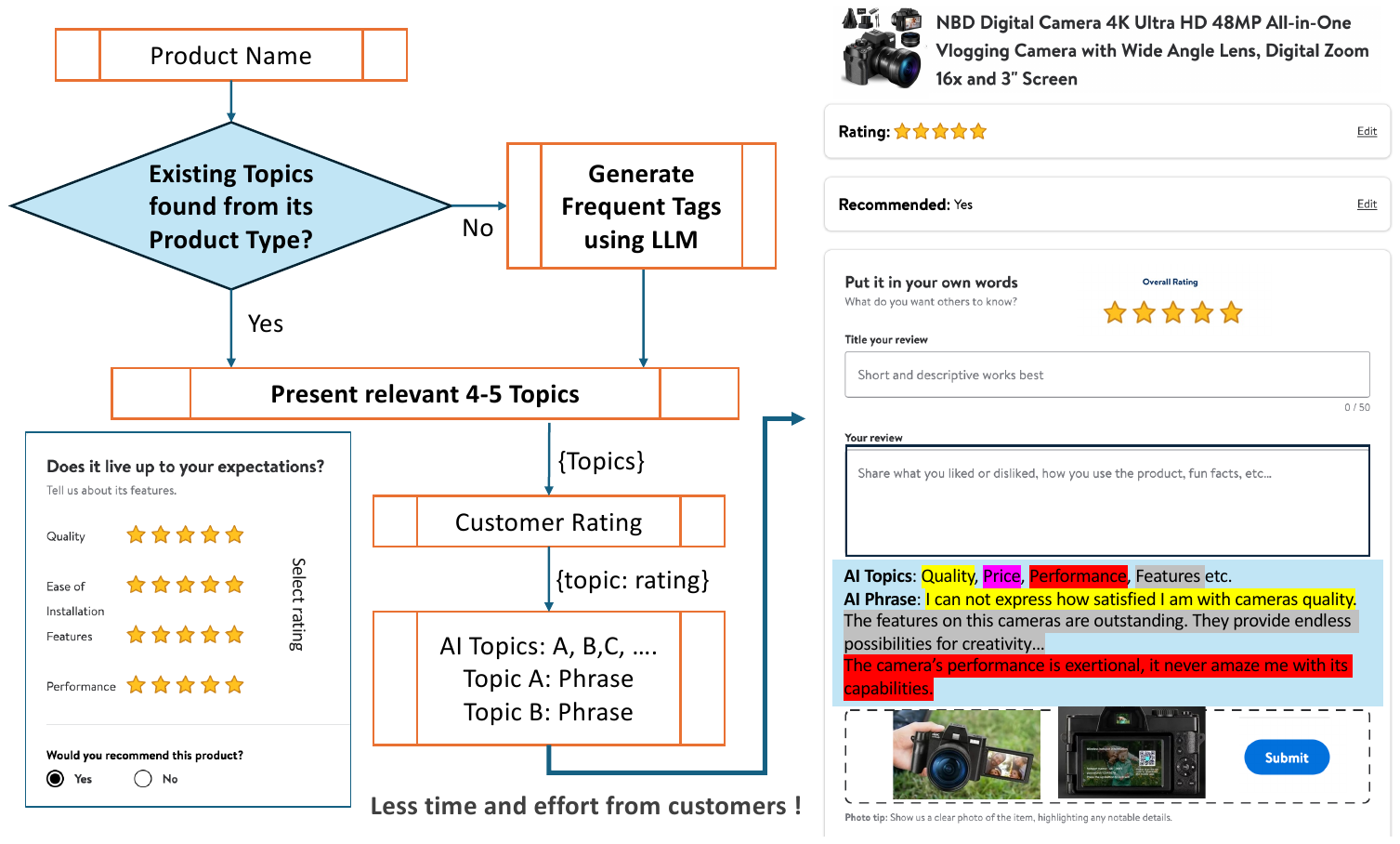}
		\caption{\bf Overview of the proposed \newmethod Framework. 1) Frequent Mentions or Topic data collection for all product types. 2) Topic generation for a new product. 3) Phrase suggestion using selected topic and rating provided by customer of selected topic.} 
		\label{fig:introimage}
	\end{center}
\vspace{-0.2in}
\end{figure*}

\hide{\textit{\textcolor{blue}{Informal Problem:} Is it possible to create an unsupervised model that necessitates minimal human annotation? Additionally, can we develop models that can suggest ideas or topics based on the given product and its type, contextual information like product name, customer's rating for the selected topics? Also, for new products can it suggest related topics from previous published reviews for similar product type? }}
When writing a product review, human perception plays a significant role in shaping the evaluation of the product. It involves the subjective interpretation of various factors such as product quality, functionality, aesthetics, and user experience. Human sentiments influence how individuals perceive the product's strengths, weaknesses, and overall value. Emotions, personal preferences, and prior experiences also impact the review process, leading to subjective judgments. Additionally, cognitive biases can influence perception, affecting the objectivity of the review. Consider the process of generating review texts: typically, we already hold sentiment polarities regarding product aspects before expressing ourselves in speech or writing. Inspired by this and different from existing work, we focus on the study of controllable topics (for new product) and phrase suggestions from the data which consists of selected topic and customer's sentiment scores. First, we collect the frequent mentions or topics for each product type and by using LLM model, we suggest a phrase based on product type and customer's sentiments regarding the given topic. The contributions of our paper are as follows:
\begin{itemize}
\item \textbf{Novel Algorithm}: We propose a novel method of automatic topics (for new product) and phrase suggestion namely \newmethod which helps in preserving sentiment so that each review will have diverse text. See Figure \ref{fig:introimage}
\item \textbf{Efficient Framework}: We propose an efficient sub-method to suggest topics for a product which does not have previous reviews. This makes our method's usability very high for a new product which has just been published on website. Our proposed framework only requires the product type of the purchased item and optionally reviews related to other items falling under the same product types when generating the reviews for new products.
\item \textbf{Case Studies}:  In this paper, we provide case studies on the $3$ top performing product types in Walmart \cite{walmart} and show the capability of our proposed framework.
\end{itemize}
The remainder of the paper is organized as follows: The problem formulation is given
in Section \ref{sec:problem}. In Section \ref{sec:method} we describe our proposed method \newmethod in detail with examples. Finally, we discuss the case studies in section  \ref{obtd:experiments} and section  \ref{sec:conclusions} concludes the paper.
\section{Problem Definition}
\label{sec:problem}
In this section, first we define the necessary terms which we are using throughout the paper and then formally define our problem. Due to Walmart Privacy Requirements, models and datasets are not open to the public. We have elaborated the details of each method in section \ref{sec:method}, and you can use LLM model of your choice.
\subsection{Basic Definitions}
\label{sec:Definitions}
\textbf{Product Type}: This refers to the category or kind of product that is being reviewed. For example, the product shown in Figure \ref{fig:exampleproblem} is "Cordless Vacuum Cleaner" and belongs to the product type "Stick Vacuums".

\textbf{Frequent Mentions or Topics}: These are common words, features, or issues that come up frequently in existing reviews of the product. 

\textbf{Topics Score}: This is the emotional tone or sentiment that is associated with each frequent mention or topic. 

\textbf{Product Name}: This is additional information about the product, such as its features, brand etc, provided by the manufacturer or seller on the website. 

\textbf{Prompt}: Text input to generate model response. Prompts can include preamble, questions, suggestions, instructions, or examples.

\subsection{Problem Statement}
Our problem can be divided into two criteria. First, when the new product is purchased and we do not have reviews for it. In that case, out framework suggests topics and phrase (positive, negative and neutral) using \newmethod. Second, in case product reviews are already available, then review system asks customers to rate the topic. Once customer rates the topic, \newmethod suggest phrase based on given rating. For both the cases, \newmethod start detecting topics in customer's written text and tagging them to its sentence . We formally define our prompt as follows:
\begin{mdframed}[backgroundcolor= blue!10,linewidth=1.3pt,] 
\textbf{Opening Prompt}: Act like a customer who purchased the product of given product type. After using the product, you  want to provide review on the website. Now you have information on product type, product related topics along with rating for each topic. \\
\textbf {Ask}: Suggest topics and respective phrase (minimum $20$ words) for selected and rated topic. \\
\textbf{Closing Prompt}: Do not mention any rating in the review text. Also, you can use synonyms for tags when generating the phrase. 
\end{mdframed}

\section{CPR Workflow and its Performance}
\label{sec:method}
In this work, we tackle the  problem in 2 steps. First, we collect the review data from Walmart.com and collect information regarding frequent mentions for the items under each product type. Next, we consolidate all frequent mentions per product type. Second, using the LLM method, we generate frequent mention by giving model input as product type, product name (optional), topics and its sentiment value or rating. Finally, we capture the sentiment score of written reviews to provide an overall rating of the product. Below we provide a detailed description of our process.

\subsection{Frequent Mentions or Topic Data Collection}
\begin{figure}
	\begin{center}
	    \includegraphics[clip,trim=0cm 0cm 0cm 0cm,width=0.45\textwidth]{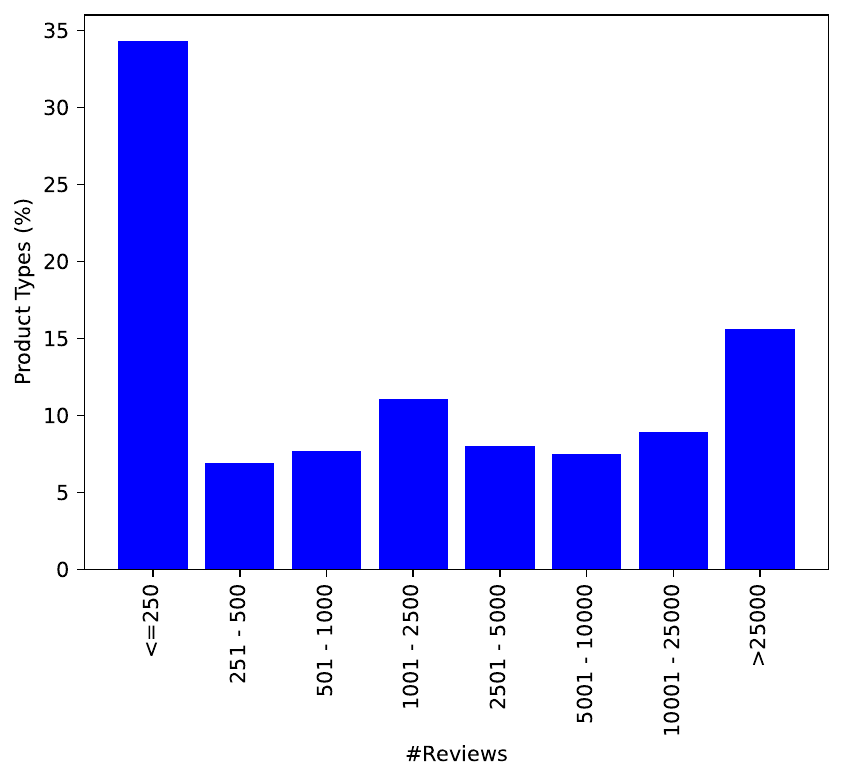}
		\caption{Product type vs number of reviews distribution.} 
		\label{fig:reviewdistribution}
	\end{center}
 \vspace{-0.2in}
\end{figure}

There are 5k+ product types available and published on Walmart.com. Currently, approximately $65.7\%$ product types have more than $250$ reviews and have frequent mentions, which is sufficient to get frequent tags for the respective product type. Figure \ref{fig:reviewdistribution} shows the distribution of reviews available per product type. For the product type \cite{Gujralpae2024} which does not have any reviews, we started by finding similar product types to generate frequent mentions or topics. Here, we explored three different methods, namely Levenshtein-based similarity, text representation (cosine similarity) and Bison-based LLM models. Levenshtein-based similarity, also known as edit distance, is a measure of the difference between two strings. It calculates the minimum number of single-character edits required to transform one string into another. The edits can be deletions, insertions, or substitutions. For example, the levenshtein distance between the words "kitten" and "sitting" is 3, as it takes three edits (substitute "k" with "s", substitute "e" with "i", and insert "g" at the end) to transform one word into the other.  We observe that Levenshtein similarity does not perform well as it doesn't consider the semantic meaning of words: Levenshtein similarity only measures the difference in terms of character-level edits. It doesn't consider the context or meaning of the words involved. For example, the product type “3D Glasses” and “Wine Glasses” have a similarity score of $0.67$. However, both product types have different contexts. We report the accuracy of the method for the Levenshtein similarity method with a threshold value of $0.5$ in Table \ref{tbl:simialrity_check}.  

Next, we explore cosine similarity where we created embeddings of each product type and computed cosine similarity w.r.t. product types belonging to the same department. It performed better than the Levenshtein similarity but was not able to capture complex relationships between embeddings. For example, it struggles to differentiate between antonyms or words with multiple meanings. It is a simplistic measure that only captures the angle between vectors, which may not adequately represent the nuances of the underlying data. Table \ref{tbl:simialrity_check} shows the performance of levenshtein and cosine similarity. 

\begin{table}[t]
	\centering
	\small
	\begin{tabular}{|c||c|c|c|c|}
	\cline{1-5}
    {\bf Method}	& {\bf $\#TPTs$} & {\bf $\#CPTs$}& {\bf $\#MPTs$} & {\bf $A(\%)$} \\	\hline
	Levenshtein Based&$410$&$186$&$89$&$45.4\%$\\\hline
         Cosine Similarity&$410$&$318$&$82$&$77.5\%$\\\hline
         LLM Based&$410$&$342$&$40$&$83.4\%$\\\hline
   \hline
	\end{tabular}
	\caption{Similar Product Type Detection Accuracy. In this we have $41$ source product types that has missing topics. For each method, we expect to get $10$ similar product types. Here, $\#TPTs$ represents total number similar product types detected, $\#CPTs$ represents number correct product types detected, $\#MPTs$ represents number of missing product types. Here $A(\%)$ represents Accuracy of the method.}
\label{tbl:simialrity_check} 
\end{table}
Further, we explored using a LLM model to find similar product types and then given a list of product types with missing topics along with similar product types found by the LLM, we generated frequent mentions or topics. We observe that topics generated by the LLM model are more accurate than traditional methods as shown in table \ref{tbl:simialrity_check}. Another limitation of traditional methods are that both methods are time-consuming and do not understand the contextual nature of product type and nuances of the underlying data. Hence, we are moving forward with the LLM model to create topics for those product types which do not have sufficient topics or reviews. Below table \ref{tbl:fmexamples} shows examples of 5 product types and its top 10 frequent mentions for reader's reference. 

\begin{table}[t]
	\centering
	\footnotesize
	\begin{tabular}{|c||c|c|}
	\cline{1-3}
    {\bf Product}	& {\bf Topics} & {\bf  Frequent Mentions} \\ 
    {\bf Type}	& {\bf via LLM} & {\bf  OR Topics } \\	\hline
	          &  &Comfort, Softness, Fit, Material, \\
       Pajamas  & No &Quality,  Price, Color, Style,\\
                &  &Design, Appearance\\\hline
                
                       &&Picture, Quality, Price, Sound, \\
     Televisions &No &Ease Of Installation,  Screen, Color, \\
                    & &Remote, Accessories, Connection\\\hline  
                    
              &&Sensitivity, Wavelength range, Software,\\
     Spectrometers &Yes &Ease of use, Versatility, Reliability, Price,\\
                   & &Customer service, Quality, Battery Life\\\hline
                   
                 &&Sturdiness, Durability,Strength\\
     Garbage Bags&No &Smell, Leak, Price, Size \\
                 & &Ease Of Use, Material, Tie\\\hline
         
          Beauty  && Gentle, Smooth, Natural, Feel, \\
      Supply Making &Yes &Hydrating, Brightening, Smell,\\
     Exfoliants & &Moisturizing, Consistency, Irritation\\\hline
   \hline
	\end{tabular}
	\caption{The five product types and their top 10 topics.}
\label{tbl:fmexamples} 
\end{table}

\subsection{Phrase Suggestion for Topics}
Once we collect the product type and frequent mention or topic data, the next step is to generate the AI phrase for the customer. Our problem is divided into two parts, first, generating a phrase of a maximum of $25$ words (or under $150$ tokens) based on given topic $T$ (mandatory), product name $P$(optional), topic's $t_i^N$ rating $r_i^N$(mandatory). In order to solve this problem, first, in section \ref{sec:bisonmethod} we describe our observation with Bison based LLM model to generate phrase. Finally, in \ref{sec:llammamethod}, we introduce \newmethod, our unified algorithm for automatic phrase generation with minimal user intervention. 

\subsubsection{Bison based LLM Model}
\label{sec:bisonmethod}
We leverage \textit{Prompt Engineering} in our study. Prompt Engineering encompasses the tasks of crafting, fine-tuning, and perfecting input prompts to clearly convey the user's intentions to a language model \cite{ekin2023prompt,chowdhery2023palm}. This methodology is vital for eliciting precise, pertinent, and cohesive responses from the model. With the ongoing evolution of language models, adept prompt engineering has emerged as a fundamental skill for users seeking to unlock the complete capabilities of LLM models \cite{anil2023palm,besta2024graph} and attain optimal outcomes across diverse applications. 

When using prompts, the challenges arise from the unclear and intricate nature of questions. Phrases used as input can have multiple meanings. For example, "Books on the shelf" could mean either books that are physically placed on a shelf or books that are available for purchase. Additionally, these phrases can be presented with complex sentence structures, including subordinate and relative clauses. Furthermore, the context surrounding these phrases plays a crucial role in determining their true meaning, making it challenging to accurately interpret them. To overcome this challenge, we use Bison based LLM model \cite{anil2023palm} to generate the first version of phrase of selected topic and its rating. Consider the following example to understand the outcome from Bison model.

\begin{mdframed}[backgroundcolor= blue!10,linewidth=1.3pt,] 
\textbf{Opening Prompt}: Act like a customer who purchased the product of given product type. After using the product, you  want to provide review on the website. Now you have information on product type, product related topics along with rating for each topic.\\
\textbf {Ask}:  Suggest phrase (minimum $20$ words) for selected and rated topic. , \\
\textbf{Input data}: Product Type: Camera Straps,\\ Topics and Ratings are: Feel: $2$ stars, features: $1$ stars , strap: $4$ stars, price: $2$ stars. \\
\textbf{Closing Prompt}: Do not mention any rating in the phrase text. Also, you can use synonyms for topic when generating the phrase.
\end{mdframed}
 
The expected review should have close to lower side of negative sentiment and average $2$ rating. The table \ref{tbl:perfumesallbaseline} shows the performance of Bison based model. We observe that the reviews generated by this method are very generic and human sentiment touch was not captured. We tried multiple combinations of parameters but did find satisfactory results. Hence, we changed our approach to leverage the LLM-based fine tuned model. 

\begin{table*}[t]
	\centering
	\footnotesize
	\begin{tabular}{|c||c|c|}
	\cline{1-3}
      {\bf Topic}  & {\bf  LLM Model} & {\bf  Phrase }  \\ \hline
       \multirow{15}{0.13\textwidth}{Feel: $2$ stars, \\ Features: $1$ stars\\       Strap: $4$ stars\\ Price: $2$ stars}&\multirow{5}{0.1\textwidth}{Bison based}&\multirow{5}{0.7\textwidth}{\textbf{Strap}: The camera strap is comfortable and well-made.\\
       \textbf{Features}: It lacks some features that would make it more versatile. \\
       \textbf{Price:} The price is reasonable, but the lack of features makes it a bit disappointing. \\
       \textbf{Feel}: It's a good choice for photographers who are looking for a basic strap, but those who need more features may want to look elsewhere.} \\
       &&\\
       &&\\
       &&\\
       &&\\ \cline{2-3}
     &\multirow{5}{0.1\textwidth}{Pre-trained LLM \cite{touvron2023llama}}&\multirow{5}{0.7\textwidth}{\textbf{Strap}: This camera strap is a disappointment in terms of quality and usability. It is not waterproof and the material is not durable. The strap is also difficult to use and adjust, making it a frustrating experience for photographers. \\
     \textbf{Price}: The price is also high compared to other similar products on the market. Overall, it is not a good value for the money.}\\
       &&\\
       &&\\
       &&\\
       &&\\ \cline{2-3}
       
       &\multirow{5}{0.1\textwidth}{Our (Fine-tuned LLM) }  &\multirow{5}{0.7\textwidth}{\textbf{Feel}: The quality of the strap left much to be desired; the material felt cheap and uncomfortable.\\
\textbf{Features}: The functionalities were shockingly basic, offering almost no practical utility for a photographer.\\
\textbf{Strap}: The strap itself was adequately strong and supportive.\\
\textbf{Price}: The cost felt disproportionately high, offering poor value for money.}\\
       &&\\
       &&\\
       &&\\
       &&\\\hline
     
               \hline
	\end{tabular}
	\caption{Performance on LLM models.}
\label{tbl:perfumesallbaseline} 
\end{table*}

\subsubsection{Fine-tuned LLM: \newmethod}
\label{sec:llammamethod}
To generate the phrases based on topic and its rating, we employ pretrained LLM \cite{touvron2023llama} based model. The prompts are meticulously crafted and fed into LLM, which in turn generates responses containing phrases. A cleaning process follows to eliminate hallucinations and incorrect format responses. 

$$phrase = LLM_{ask}(Prompt(PT, T_i^N, S_i^N))$$

In the process delineated, $PT$ is the product type, $T$ are topic, $S$ are sentiment values and $phrase$ signifies the phrase generated by the model. The function Prompt is tasked with the guided prompts that are instrumental in the review generation workflow. The term LLM $ask$ designates the phrase generation phase of the LLM system, which processes the prompts to procure a preliminary phrase text.  The table \ref{tbl:perfumesallbaseline} shows the outcome generated by the pre-trained LLM model. It added additional context in the phrase which was never asked in the prompt and might be not suitable for the customer. For example, during phrase generation, "waterproof" context was not provided in prompt and not described by customer, but model added this additional content to the review. Here, we aim to generate the phrases based on content provided by customers. Hence, we observe that fine-tuning of model is required.  

We adopt most of the pre-training setting and model architecture from LLM \cite{touvron2023llama} (same standard transformer architecture, pre-normalization using RMSNorm, SwiGLU activation function, and rotary positional embeddings). We fine-tuned LLM model on a NVIDIA GPU. We fine-tune our base model using a smaller dataset consisting $12K$ reviews and approximate $200$ reviews per product type and processed the data to match the LLM prompt format. 

The components of training data used to fine-tune the model are:
\begin{itemize}
    \item "instruction" : This includes the prompt instruction to be followed
    \item "context" : Represents the format of product type, tags of frequent mentions along with star rating for the reviews
    \item "response" : This has the final output format of title and summarized review used to fine-tune the LLM model. 
\end{itemize}
We also enable gradient check pointing to reduce memory usage during fine-tuning of our proposed model. Next, we create $4$-bit quantization with $NF4$ type configuration using BitsAndBytes from the transformers python library. In general, fine-tuning of pre-trained LLM requires updating all of the model's parameters, which is computationally expensive and requires massive amounts of data. But, we use Parameter-Efficient Fine-Tuning (PEFT) \cite{peft} for updating a small subset of the model's most influential parameters. We specifically used Low-Rank Adaptation (LoRA) \cite{hu2021lora} that decomposes a large matrix into two smaller low-rank matrices in the attention layers. We use modules like "q\_proj", "o\_proj" to apply the LoRA update matrices. This drastically reduces the number of parameters that need to be fine-tuned. To set up the tokenizer, we added padding on the left to make training use less memory. By leveraging Hugging Face libraries like transformers, accelerate, PEFT, TRL, and BitsAndBytes, we were able to successfully fine-tune the LLM model on a consumer GPU.

The table  \ref{tbl:perfumesallbaseline} and section \ref{obtd:experiments} shows the performance of fine-tuned LLM model. Let's break down the sentiment in the review generated by \textit{Bison model}, it has $3$ positive aspects, namely "comfortable and well-made", "price is reasonable", "good choice for photographers who are looking for a basic strap" and $3$ negative aspects namely "lacks some features that would make it more versatile", "lack of features makes it a bit disappointing", "those who need more features may want to look elsewhere". Based on sentiment analysis \cite{loureiro2022timelms}, the compound score is $0.6$, which translates to $3$ stars ratings. This over-evaluates the customer purchase. The average rounded rating for given tags is $2$ stars which leans towards negative reviews. Similarly, reviews generated by the \textit{pre-trained LLM model} have mostly negative aspects, namely "disappointment in terms of quality and usability", "not waterproof", "material is not durable", "difficult to use and adjust", "frustrating experience for photographers", "price is also high compared to other similar products", "not a good value for the money" and translate to sentiment score as $0.2$ and rating $1$ star. Also, it misses suggesting other topics. This under-evaluates the customer purchase. However, the fine-tuned LLM model \textit{\newmethod} is able to capture the right mixture of customer's sentiment. The negative aspects captured by the model are "the material felt cheap and uncomfortable", "almost no practical utility", "doesn't provide the desired level of comfort and support", "cost felt disproportionately high", "offering poor value for money" and the neutral/positive aspect is "strap itself was adequately strong and supportive". The sentiment score is $0.3$ which translates to $2$ stars rating. We tested $10k$ reviews and found out that behavior is consistent with most of the reviews.

\subsection{Evaluating the Phrase Quality}
We evaluate the outcome with BLEU (Bilingual Evaluation Understudy) \cite{papineni2002bleu} score. The BLEU score is a number between $0$ and $1$ that measures the similarity of the machine-translated text to a set of high-quality reference translations. The scores are calculated for individual translated segments—generally sentences—by comparing them with a set of good-quality reference translations. These scores are then averaged over the whole corpus to reach an estimate of the translation’s overall quality. Neither intelligibility nor grammatical correctness are taken into account. 
To compute the BLEU score, we tokenized all the phrase text of the product and removed stop words. This product consists of $2.3k+$ reviews and $57\%$ reviews are eligible to compute BLEU score as the reference and candidates require the same word length. The table \ref{tbl:BLEUexample} shows the example of BLEU scores generated by all three methods. High BLEU score means many $n$-grams in the hypothesis texts meet the gold-standard references. \begin{table}[t]
	\centering
	\footnotesize
	\begin{tabular}{|c||c|c|c|}
	\cline{1-4}
      {\bf  n-gram } &{\bf Bison based}&{\bf Pre-trained LLM}& {\bf  \newmethod }   \\ \hline
       Cumulative 1-gram &$0.693$&$0.419$& $0.778$\\ \hline
       Cumulative 2-gram &$0.371$&$0.199$ & $0.497$\\ \hline
       Cumulative 3-gram  &$0.207$&$0.161$& $0.294$\\ \hline
       Cumulative 4-gram &$0.132$&$0.123$ & $0.168$\\ \hline
 \hline
	\end{tabular}
	\caption{Average BLEU score for phrase suggested in table \ref{tbl:perfumesallbaseline}.The score $>0.4$ reflects high quality understandable, adequate, and fluent translations \cite{bleuscoreggle}}
\label{tbl:BLEUexample} 
\end{table}
\section{Case Studies}
\label{obtd:experiments}
Rather than doing traditional experiments, we provide case studies for $3$ different product types namely Toys, Apparel and Perfumes. Here we answer following two research questions:
\begin{itemize}
\item \textbf{Q1:} Does \newmethod able to provide relevant additional topics based on given product type?
\item \textbf{Q2:} Does \newmethod able to generate phrase based on given rating effectively for different product types?
\end{itemize}
\begin{table*}[t]
	\centering
	\footnotesize
	\begin{tabular}{|c||c|c|c|c|c|}
	\cline{1-6}
     {\bf  Product } & \multirow{2}{*}{\bf Topics}  & {\bf  Phrase }  & {\bf  Sent. }  & {\bf  BLEU } & {\bf Final}\\ 
     {\bf  Type } &  & {\bf  Suggestion }  & {\bf  Score }  & {\bf  Score } & {\bf Rating}\\ \hline
 & \multirow{15}{0.08\textwidth}{Smell: 1 stars\\ Price: 2 stars\\ Warm: 2 star \\ Long Lasting: 1 star}   & \multirow{5}{0.56\textwidth}{\textbf{Smell}: The fragrance was quite unpleasant; it didn't align with my expectations at all. \\ \textbf{Price}: For the quality, the cost felt excessive and not justified. \\ \textbf{Warm}:"It lacked the cozy, inviting feel I typically look for in a scent  \\ \textbf{Long Lasting}: The scent disappeared almost immediately after application. }  &\multirow{5}{0.03\textwidth}{$0.17$}&\multirow{5}{0.03\textwidth}{$0.52$}&\multirow{5}{0.03\textwidth}{$1.0$}\\ 
              &&&&&\\  
               &&&&&\\  
                &&&&&\\  
              &&&&&\\ \cline{3-6} 
         &     & \multirow{5}{0.56\textwidth}{\textbf{Smell}: The scent was rather off-putting and not at all what I hoped for. \\ \textbf{Price}: The value for the money was poor; it seemed overpriced for what it offered. \\ \textbf{Warm}: The fragrance didn't have the comforting warmth that I expected. \\ \textbf{Long Lasting}: It faded incredibly quickly; the longevity was extremely poor.} &\multirow{5}{0.03\textwidth}{$0.16$}&\multirow{5}{0.03\textwidth}{$0.46$}&\multirow{5}{0.03\textwidth}{$1.5$}\\  
       \textbf{Perfumes}        &&&&&\\ 
         &&&&&\\  
                &&&&&\\ 
              &&&&&\\ \cline{3-6} 
        &     & \multirow{5}{0.56\textwidth}{\textbf{Smell}:The aroma was disappointing; it lacked any appealing notes  \\ \textbf{Price}: The price point was too high considering the overall experience. \\ \textbf{Warm}: The scent felt cold and lacked the desired depth and warmth. \\ \textbf{Long Lasting}:The fragrance had no staying power; it was gone within minutes.}&\multirow{5}{0.03\textwidth}{$0.19$}&\multirow{5}{0.03\textwidth}{$0.44$}&\multirow{5}{0.03\textwidth}{$1.3$}\\ 
              &&&&&\\ 
                &&&&&\\  
                &&&&&\\ 
              &&&&&\\ \cline{3-6} 
 \hline
 \hline
 & \multirow{15}{0.08\textwidth}{Size: 4 stars \\ Softness: 4 stars\\ Quality: 5 star \\ Carry: 5 star}   & \multirow{5}{0.56\textwidth}{\textbf{Size}: The dimensions are just right, perfect for little hands to manage. \\
 \textbf{Softness}:It has a lovely, smooth texture, very comfortable for cuddling. \\
 \textbf{Quality}: I'm really impressed with the high standard of materials and construction.\\
 \textbf{Carry}:Lightweight and convenient, this toy is perfect for travel. }  &\multirow{5}{0.03\textwidth}{$0.88$}&\multirow{5}{0.03\textwidth}{$0.51$}&\multirow{5}{0.03\textwidth}{$4.7$}\\ 
              &&&&&\\
                &&&&&\\  
                &&&&&\\ 
              &&&&&\\ \cline{3-6} 

         &     & \multirow{5}{0.56\textwidth}{\textbf{Size}:It's a good, substantial toy, definitely not too small. \\
 \textbf{Softness}: The material is quite gentle and pleasant to the touch.\\
 \textbf{Quality}: The craftsmanship is exceptional; this toy is built to last.\\
 \textbf{Carry}:It's incredibly easy to take along; the portability is fantastic.} &\multirow{5}{0.03\textwidth}{$0.83$}&\multirow{5}{0.03\textwidth}{$0.43$}&\multirow{5}{0.03\textwidth}{$4.0$}\\ 
       \textbf{Toys}        &&&&&\\  
         &&&&&\\  
                &&&&&\\ 
              &&&&&\\ \cline{3-6} 
        &  & \multirow{5}{0.56\textwidth}{\textbf{Size}: This toy has a nice, adequate size, making it easy to spot and play with.\\
 \textbf{Softness}: The toy is agreeably soft, making it safe and enjoyable for kids.\\
 \textbf{Quality}: This toy shows excellent durability and attention to detail. \\
 \textbf{Carry}:This toy is very handy to move around, making it a great companion.
}&\multirow{5}{0.03\textwidth}{$0.75$}&\multirow{5}{0.03\textwidth}{$0.44$}&\multirow{5}{0.03\textwidth}{$4.5$}\\ 
              &&&&&\\  
                &&&&&\\  
                &&&&&\\ 
              &&&&&\\ \cline{3-6} 
               \hline
 \hline
 & \multirow{15}{0.08\textwidth}{Size: 2 stars \\ Fit: 3 stars \\ Appearance: 3 star \\  color: 2 star}   & \multirow{5}{0.56\textwidth}{\textbf{Fit}:  The top fits well, but the size is a bit small.\\
 \textbf{Appearance}:   The ruffles are cute, but the color is not quite what I expected.\\
 \textbf{Color}:   The color is a bit dull and not as vibrant as I hoped.\\
\textbf{Size}:   The size is small, but the material is comfortable.
}  &\multirow{5}{0.03\textwidth}{$0.42$}&\multirow{5}{0.03\textwidth}{$0.41$}&\multirow{5}{0.03\textwidth}{$3.1$}\\ 
              &&&&&\\ 
                &&&&&\\  
                &&&&&\\ 
              &&&&&\\ \cline{3-6} 
          &     & \multirow{5}{0.56\textwidth}{\textbf{Fit}: It sat alright, not perfect, but I could make it work with some adjustments.\\
\textbf{Appearance}: The design was decent, though I expected a little more visual appeal.\\
\textbf{Color}: The hue was disappointing; it didn't complement my skin tone as I had hoped.\\
\textbf{Size}: The dimensions were off; it felt restrictive and didn't match the standard sizing I'm used to.
} &\multirow{5}{0.03\textwidth}{$0.35$}&\multirow{5}{0.03\textwidth}{$0.49$}&\multirow{5}{0.03\textwidth}{$2.8$}\\ 
       \textbf{Ruffled Tops}        &&&&&\\  
         &&&&&\\  
                &&&&&\\ 
              &&&&&\\ \cline{3-6} 

        &     & \multirow{8}{0.56\textwidth}{\textbf{Fit}: The overall shape was acceptable, though I had to fuss with it a bit to get it to look decent.\\
\textbf{Appearance}:It was an average-looking top; it didn't stand out in any particular way.\\
\textbf{Color}:The coloration was off; it looked faded and wasn't what I expected from the image.\\
\textbf{Size}: It was a tight squeeze; I was disappointed in how snug it was compared to other tops I own.}&\multirow{8}{0.03\textwidth}{$0.30$}&\multirow{8}{0.03\textwidth}{$0.42$}&\multirow{8}{0.03\textwidth}{$2.6$}\\ 
              &&&&&\\ 
              &&&&&\\ 
              &&&&&\\ 
              &&&&&\\ 
                &&&&&\\  
                &&&&&\\ 
              &&&&&\\ \cline{3-6} 
               \hline
               \hline
	\end{tabular}
	\caption{\newmethod and its performance on 3 selected product types.}
\label{tbl:perfumesall} 
\end{table*}
\subsection{Experiment Setup}
\begin{figure}
	\begin{center}
	    \includegraphics[clip,trim=0cm 1cm 0cm 1cm,width=0.47\textwidth]{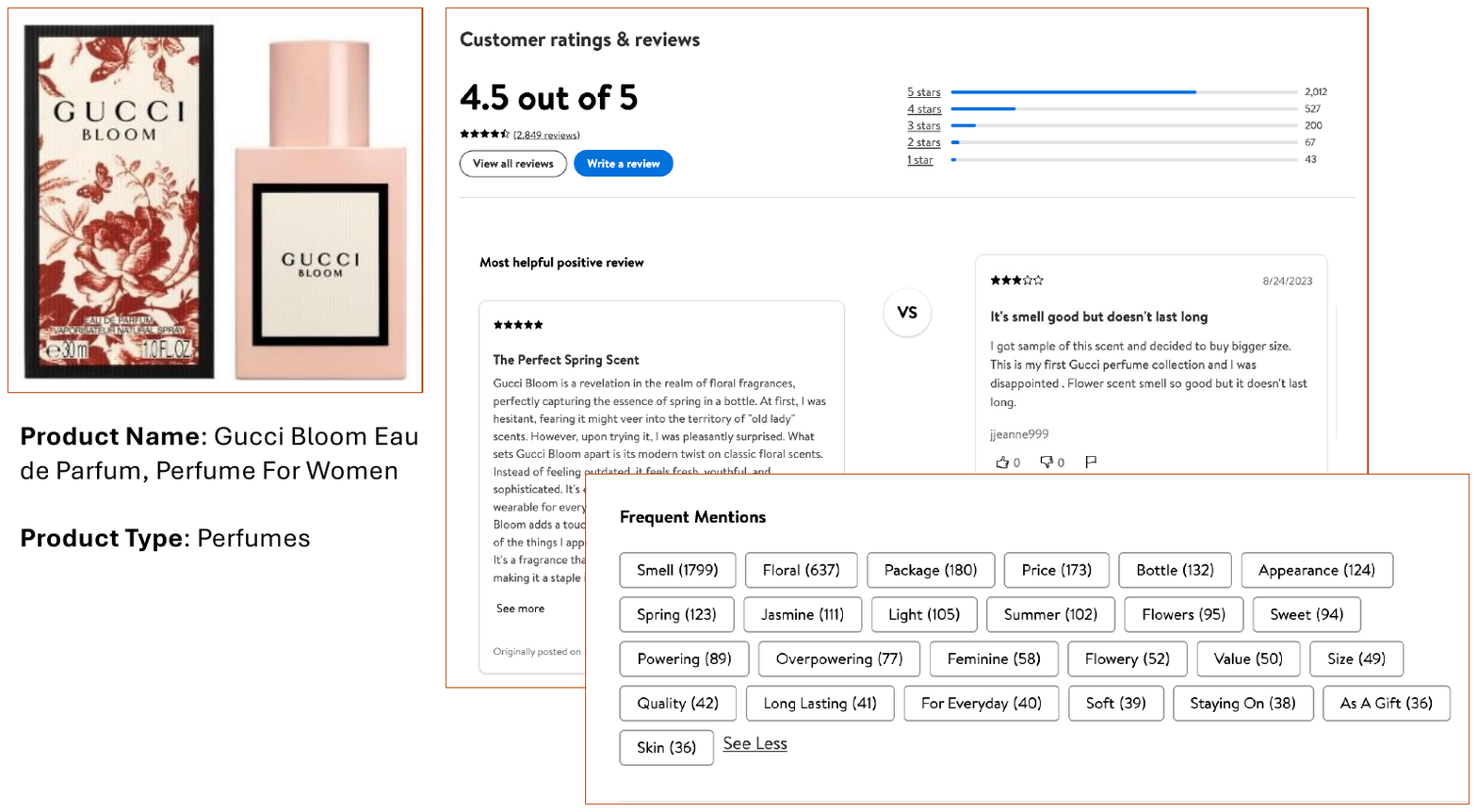}
		\caption{Example of "Perfumes" product type} 
		\label{fig:perfumes}
	\end{center}
\end{figure}

In first experiment, we ask \newmethod to generate a negative sentimental review for product type "Perfumes". In this, we selected product "Gucci Bloom Eau de Parfum, Perfume For Women" shown in Figure \ref{fig:perfumes} for our experiment. For the Perfumes product category, topics are smell, price, quality, sweet, long lasting, over powering etc. We assumed that customer did not like the product. Here, we selected 4 topics and their ratings are as "Smell": 1 stars, "Price": 2 stars, "Warm": 2 star and "Long Lasting": 1 star. 

In the second experiment, we ask \newmethod to generate positive sentimental review for product type "Stuffed toys \& animals". 
In this, we selected product "Fisher-Price Laugh \& Learn Smart Stages Puppy Plush Learning Toy" for our experiment. For this product category, topics are baby, price, as a gift, lights,learning, color, size, soft, entertainment, appearance, carry etc. We assumed that customer liked the product. Here, we selected 4 topics and their ratings are as "Size": 4 stars, "Softness": 4 stars, "Quality": 5 star and "Carry": 5 star. 

In the third experiment, we ask \newmethod to generate neutral sentimental review for product type "Ruffled Tops". 
In this, we selected product "Fantaslook Blouses for Women Floral Print V Neck Ruffle Short Sleeve Shirts Casual Summer Tops" for our experiment. For this product category, topics are fit, material, color, comfort, appearance, flattering, wash, stretch etc. We assumed that customer had mixed feelings about the product. Here, we selected 4 topics and their ratings are as "Size": 2 stars, "Fit": 3 stars, "Appearance": 3 star and "color": 2 star. 

\subsection{A1: Topic Suggestion Quality}
Here we test \newmethod capability to suggest relevant topics based on given product type. For this experiment, we choose randomly $500$ product types which have existing frequent mention or topic associated with it and another $500$ product types which does not have any reviews in the past. The accuracy is checked based on how many topics were correctly predicted for PTs where we already have topics (top 10 topics). For later set, we used product descriptions and extracted the topics from them using existing method \cite{campos2020yake} and computed the accuracy of topics. The performance is shown in table \ref{tbl:producttype_topic_check}. The proposed method \newmethod is able to suggest average $79.3\%$ of the topics for both the scenarios.
\begin{table}[t]
	\centering
	\small
	\begin{tabular}{|c||c|c|c|c|}
	\cline{1-5}
 \multirow{2}{*}{\bf Topics}&  {\bf Total} & {\bf Relevant}&  {\bf Irrelevant}   & \multirow{2}{*}{\bf Accuracy}   \\
       &  {\bf Topics} & {\bf Topics}&  {\bf Topics}  & \\ \hline
	  Available&$5,000$&$4,172$&$128$&$83.5\%$\\\hline
        Not Available&$5,000$&$3,760$&$476$&$75.2\%$\\\hline
   \hline
	\end{tabular}
	\caption{Performance of topic suggestion method of \newmethod. Topic is irrelevant if it is out of scope of the given product type.}
\label{tbl:producttype_topic_check} 
\end{table}

\subsection{A2: Phrase Suggestion Diversity}
The Table \ref{tbl:perfumesall} shows the phrase generated by our proposed model \newmethod and its evaluation using BLEU and sentiment Score. For product type "Perfumes", the phrase generated focused on the disappointment in purchase related to high price, the weak scent and an overall a under-whelming experience. Our proposed model \newmethod is able to capture this sentiment. Based on numerical evaluation, generated phrases have an average sentiment score $0.17$ and average human readability score is $0.49$ ($>0.4$ reflects high quality understandable, adequate, and fluent translations \cite{bleuscoreggle}) which reflects phrases has close to perfect overlap with the reference translations. Next, we tested our model performance on product type "Stuffed toys \& animals", the review generated summarizes the adorable nature of the toy and how it is perfect for kids; appreciating its size, quality and good craftsmanship's of stuffed toy. Notice that the phrase generated using our model, for example "Perfumes" is completely the opposite sentiment of this example. Positive reviews play a crucial role in future customer's purchasing decision and help to build trust and credibility for any retail or online business. They show that other customers have had good experiences with your products or services, which can reassure potential customers that they can trust your business. Here, the average sentiment score is $0.82$ and human readability score is high as well. Overall, customers can chose to select the rating between $4$ to $5$. Last but not least, we tested on product type "Ruffled Tops", we want to capture the neutral sentiment of customer purchase. The generated phrases summarize that the top has indecisive look and highlighted that it is pretty but the fit is not as expected. This kind of phrase suggests that the customer has neither a positive nor negative opinion. They may be indifferent, mixed, or simply not expressing any particular emotion.  Our model is able to successfully balance out the pros and cons mentioned in the form of topics and sentiment scores.
\section{CONCLUSIONS AND FUTURE WORK}
\label{sec:conclusions}
In this paper, we design \newmethod for topic and phrase suggestion task. Traditional text retrieval based methods have limitations in text generation which are depended on input specific topics and sentiment scores. To overcome these obstacles, we proposed \newmethod and find that our LLM-based model with prompt engineering can produce reliable human readable phrase with high quality and  effectively capture the sentiments of customers. We presented real-time product case studies and showed how this will save time for customer.

However, we notice that there is still room for a lot of improvements in our proposed model. In future work, the focus will be on improving the naturalness and coherence of generated phrases which can make them more convincing and engaging for readers. Also, we will improve our proposed model to the personalization \cite{yan2023personalized} domain. Incorporating user-specific information or preferences into the review generation process can make the reviews more tailored and relevant to individual users, leading to a more personalized user experience.

\bibliographystyle{plain}
\bibliography{BIB/refs}

\begin{thebibliography}{10}

\bibitem{anil2023palm}
Rohan Anil, Andrew~M Dai, Orhan Firat, Melvin Johnson, Dmitry Lepikhin, Alexandre Passos, Siamak Shakeri, Emanuel Taropa, Paige Bailey, Zhifeng Chen, et~al.
\newblock Palm 2 technical report.
\newblock {\em arXiv preprint arXiv:2305.10403}, 2023.

\bibitem{besta2024graph}
Maciej Besta, Nils Blach, Ales Kubicek, Robert Gerstenberger, Michal Podstawski, Lukas Gianinazzi, Joanna Gajda, Tomasz Lehmann, Hubert Niewiadomski, Piotr Nyczyk, et~al.
\newblock Graph of thoughts: Solving elaborate problems with large language models.
\newblock In {\em Proceedings of the AAAI Conference on Artificial Intelligence}, volume~38, pages 17682--17690, 2024.

\bibitem{walmart}
Walmart Business.
\newblock {Walmart.com}.
\newblock \url{https://www.walmart.com/}.
\newblock [2024 Walmart. All Rights Reserved.].

\bibitem{campos2020yake}
Ricardo Campos, V{\'\i}tor Mangaravite, Arian Pasquali, Al{\'\i}pio Jorge, C{\'e}lia Nunes, and Adam Jatowt.
\newblock Yake! keyword extraction from single documents using multiple local features.
\newblock {\em Information Sciences}, 509:257--289, 2020.

\bibitem{carlson2023complementing}
Keith Carlson, Praveen~K Kopalle, Allen Riddell, Daniel Rockmore, and Prasad Vana.
\newblock Complementing human effort in online reviews: A deep learning approach to automatic content generation and review synthesis.
\newblock {\em International Journal of Research in Marketing}, 40(1):54--74, 2023.

\bibitem{celikyilmaz2020evaluation}
Asli Celikyilmaz, Elizabeth Clark, and Jianfeng Gao.
\newblock Evaluation of text generation: A survey.
\newblock {\em arXiv preprint arXiv:2006.14799}, 2020.

\bibitem{chakraborty2022attribute}
Ishita Chakraborty, Minkyung Kim, and K~Sudhir.
\newblock Attribute sentiment scoring with online text reviews: Accounting for language structure and missing attributes.
\newblock {\em Journal of Marketing Research}, 59(3):600--622, 2022.

\bibitem{chevalier2018channels}
Judith~A Chevalier, Yaniv Dover, and Dina Mayzlin.
\newblock Channels of impact: User reviews when quality is dynamic and managers respond.
\newblock {\em Marketing Science}, 37(5):688--709, 2018.

\bibitem{chintagunta2010effects}
Pradeep~K Chintagunta, Shyam Gopinath, and Sriram Venkataraman.
\newblock The effects of online user reviews on movie box office performance: Accounting for sequential rollout and aggregation across local markets.
\newblock {\em Marketing science}, 29(5):944--957, 2010.

\bibitem{chowdhery2023palm}
Aakanksha Chowdhery, Sharan Narang, Jacob Devlin, Maarten Bosma, Gaurav Mishra, Adam Roberts, Paul Barham, Hyung~Won Chung, Charles Sutton, Sebastian Gehrmann, et~al.
\newblock Palm: Scaling language modeling with pathways.
\newblock {\em Journal of Machine Learning Research}, 24(240):1--113, 2023.

\bibitem{costa2018automatic}
Felipe Costa, Sixun Ouyang, Peter Dolog, and Aonghus Lawlor.
\newblock Automatic generation of natural language explanations.
\newblock In {\em Proceedings of the 23rd international conference on intelligent user interfaces companion}, pages 1--2, 2018.

\bibitem{ekin2023prompt}
Sabit Ekin.
\newblock Prompt engineering for chatgpt: a quick guide to techniques, tips, and best practices.
\newblock {\em Authorea Preprints}, 2023.

\bibitem{fan2022automatic}
Xiaochuan Fan, Chi Zhang, Yong Yang, Yue Shang, Xueying Zhang, Zhen He, Yun Xiao, Bo~Long, and Lingfei Wu.
\newblock Automatic generation of product-image sequence in e-commerce.
\newblock In {\em Proceedings of the 28th ACM SIGKDD Conference on Knowledge Discovery and Data Mining}, pages 2851--2859, 2022.

\bibitem{guo2021conditional}
Bin Guo, Hao Wang, Yasan Ding, Wei Wu, Shaoyang Hao, Yueqi Sun, and Zhiwen Yu.
\newblock Conditional text generation for harmonious human-machine interaction.
\newblock {\em ACM Transactions on Intelligent Systems and Technology (TIST)}, 12(2):1--50, 2021.

\bibitem{hu2021lora}
Edward~J Hu, Yelong Shen, Phillip Wallis, Zeyuan Allen-Zhu, Yuanzhi Li, Shean Wang, Lu~Wang, and Weizhu Chen.
\newblock Lora: Low-rank adaptation of large language models.
\newblock {\em arXiv preprint arXiv:2106.09685}, 2021.

\bibitem{korhonen2019proceedings}
Anna Korhonen, David Traum, and Llu{\'\i}s M{\`a}rquez.
\newblock Proceedings of the 57th annual meeting of the association for computational linguistics.
\newblock In {\em Proceedings of the 57th Annual Meeting of the Association for Computational Linguistics}, 2019.

\bibitem{bleuscoreggle}
Alon Lavie.
\newblock Evaluating the output of machine translation systems.
\newblock \url{https://www.cs.cmu.edu/%7Ealavie/Presentations/MT-Evaluation-MT-Summit-Tutorial-19Sep11.pdf}, 2011.
\newblock 13th MT Summit Tutorial,Xiamen, China.

\bibitem{loureiro2022timelms}
Daniel Loureiro, Francesco Barbieri, Leonardo Neves, Luis~Espinosa Anke, and Jose Camacho-Collados.
\newblock Timelms: Diachronic language models from twitter.
\newblock {\em arXiv preprint arXiv:2202.03829}, 2022.

\bibitem{ludwig2013more}
Stephan Ludwig, Ko~De~Ruyter, Mike Friedman, Elisabeth~C Br{\"u}ggen, Martin Wetzels, and Gerard Pfann.
\newblock More than words: The influence of affective content and linguistic style matches in online reviews on conversion rates.
\newblock {\em Journal of marketing}, 77(1):87--103, 2013.

\bibitem{peft}
Sourab Mangrulkar, Sylvain Gugger, Lysandre Debut, Younes Belkada, Sayak Paul, and Benjamin Bossan.
\newblock Peft: State-of-the-art parameter-efficient fine-tuning methods.
\newblock \url{https://github.com/huggingface/peft}, 2022.

\bibitem{ftcfakereview2023}
Office of~Public~Affairs.
\newblock Federal trade commission announces final rule banning fake reviews and testimonials.
\newblock https://www.ftc.gov/news-events/news/press-releases/2024/08/federal-trade-commission-announces-final-rule-banning-fake-reviews- testimonials, 2023.
\newblock [Press Releases, June 30, 2023].

\bibitem{papineni2002bleu}
Kishore Papineni, Salim Roukos, Todd Ward, and Wei-Jing Zhu.
\newblock Bleu: a method for automatic evaluation of machine translation.
\newblock In {\em Proceedings of the 40th annual meeting of the Association for Computational Linguistics}, pages 311--318, 2002.

\bibitem{perez2023efficiency}
Amparo Perez-Castro, Mar{\'\i}a del~Roc{\'\i}o Mart{\'\i}nez-Torres, and SL~Toral.
\newblock Efficiency of automatic text generators for online review content generation.
\newblock {\em Technological Forecasting and Social Change}, 189:122380, 2023.

\bibitem{ruder2016hierarchical}
Sebastian Ruder, Parsa Ghaffari, and John~G Breslin.
\newblock A hierarchical model of reviews for aspect-based sentiment analysis.
\newblock {\em arXiv preprint arXiv:1609.02745}, 2016.

\bibitem{Gujralpae2024}
Apurva Sinha and Ekta Gujral.
\newblock Pae: Llm-based product attribute extraction for e-commerce fashion trends.
\newblock {\em arXiv preprint arXiv:2405.17533}, 2024.

\bibitem{tai2018automatic}
Yu~Tai, Hui He, WeiZhe Zhang, and Yanguo Jia.
\newblock Automatic generation of review content in specific domain of social network based on rnn.
\newblock In {\em 2018 IEEE Third International Conference on Data Science in Cyberspace (DSC)}, pages 601--608. IEEE, 2018.

\bibitem{touvron2023llama}
Hugo Touvron, Louis Martin, Kevin Stone, Peter Albert, Amjad Almahairi, Yasmine Babaei, Nikolay Bashlykov, Soumya Batra, Prajjwal Bhargava, Shruti Bhosale, et~al.
\newblock Llama 2: Open foundation and fine-tuned chat models.
\newblock {\em arXiv preprint arXiv:2307.09288}, 2023.

\bibitem{voss1980text}
James~F Voss, Gregg~T Vesonder, and George~J Spilich.
\newblock Text generation and recall by high-knowledge and low-knowledge individuals.
\newblock {\em Journal of verbal Learning and verbal Behavior}, 19(6):651--667, 1980.

\bibitem{yan2023personalized}
An~Yan, Zhankui He, Jiacheng Li, Tianyang Zhang, and Julian McAuley.
\newblock Personalized showcases: Generating multi-modal explanations for recommendations.
\newblock In {\em Proceedings of the 46th International ACM SIGIR Conference on Research and Development in Information Retrieval}, pages 2251--2255, 2023.

\bibitem{zang2017towards}
Hongyu Zang and Xiaojun Wan.
\newblock Towards automatic generation of product reviews from aspect-sentiment scores.
\newblock In {\em Proceedings of the 10th International Conference on Natural Language Generation}, pages 168--177, 2017.

\bibitem{zhao2019review}
Lujun Zhao, Kaisong Song, Changlong Sun, Qi~Zhang, Xuanjing Huang, and Xiaozhong Liu.
\newblock Review response generation in e-commerce platforms with external product information.
\newblock In {\em The world wide web conference}, pages 2425--2435, 2019.

\end{thebibliography}
\end{document}